\documentclass[12pt,twocloumn]{article}
\usepackage{amssymb}
\usepackage{dcolumn}

\usepackage[dvips]{epsfig}

\newcommand{\be}{\begin{equation}}
\newcommand{\ee}{\end{equation}}
\def\bea{\begin{eqnarray}}
\def\eea{\end{eqnarray}}

\newcommand{\bn}{\begin{eqnarray}}
\newcommand{\en}{\end{eqnarray}}

\newcommand{\p}{\partial}

\newcommand{\nn}{\nonumber}

\newcommand{\no}{\noindent}

\newcommand{\ttf}{\tilde{f}}

\def\bea{\begin{eqnarray}}
\def\eea{\end{eqnarray}}

\newcommand{\beq}{\begin{eqnarray}}
\newcommand{\eeq}{\end{eqnarray}}
\begin{document}

\title{\textbf{Massive ``spin-2'' theories in arbitrary $D \ge 3$ dimensions}}
\author{D. Dalmazi\footnote{dalmazi@feg.unesp.br},
A.L.R. dos Santos\footnote{alessandroribeiros@yahoo.com.br}, E.L. Mendon\c ca\footnote{elias.fis@gmail.com} \\
\textit{{UNESP - Campus de Guaratinguet\'a - DFQ} }\\
\textit{{Avenida Dr. Ariberto Pereira da Cunha, 333} }\\
\textit{{CEP 12516-410 - Guaratinguet\'a - SP - Brazil.} }\\}
\date{\today}
\maketitle

\begin{abstract}

Here we show that in arbitrary dimensions $D\ge 3$ there are two families of second order Lagrangians describing
massive ``spin-2'' particles via a nonsymmetric rank-2 tensor. They differ  from the usual Fierz-Pauli theory in
general. At zero mass one of the families is Weyl invariant. Such massless theory has no particle content in
$D=3$ and gives rise, via master action, to a dual higher order ( in derivatives ) description of massive spin-2
particles in $D=3$ where both the second and the fourth order terms are Weyl invariant, contrary to the
linearized New Massive Gravity. However, only the fourth order term is invariant under arbitrary antisymmetric
shifts. Consequently, the antisymmetric part of the tensor $e_{[\mu\nu]}$ propagates at large momentum as
$1/p^2$ instead of $1/p^4$. So, the same kind of obstacle for the renormalizability of the New Massive Gravity
reappears in this nonsymmetric higher order description of massive spin-2 particles.

\end{abstract}

\newpage

\section{ Introduction}

In the last years there has been a growing interest in massive gravity
theories, see the review works \cite{hinter,rham} and references therein. From
the theoretical point of view, much of the recent interest in the subject has
been triggered by the ghost free model of \cite{rg,rgt}, see also \cite{hr1},
both in $D=4$. Recent developments in $D=3$ on the other hand, were partially
motivated by the New Massive Gravity (NMG) of \cite{bht}. All those works have
the Fierz-Pauli (FP) model \cite{fp} of massive spin-2 particles as underlying
free theory. The model of \cite{rg,rgt} reduces to \cite{fp} at quadratic
(free) level. Although this is not true for the NMG, the linearized version of
NMG is closely connected with the FP theory, see \cite{bht} and also
\cite{sd4}.

It is desirable to look for alternative non Fierz-Pauli ( nFP) descriptions of
massive spin-2 particles and their possible nonlinear extensions, in order to
verify which physical features of recent massive gravities are model
independent. The basic field in the FP theory  is a symmetric rank-2
tensor\footnote{Throughout this work we use $\eta_{\mu\nu}=diag(-,+,\cdots ,+)$
and $e_{(\mu\nu)}= (e_{\mu\nu} + e_{\nu\mu})/2 \, , \, e_{[\mu\nu]}=
(e_{\mu\nu} - e_{\nu\mu})/2 $.} $e_{\mu\nu} = e_{(\mu\nu)}$. Recently, other
second order models for massive spin-2 particles in $D=4$ have been suggested
\cite{ms,nfp,rank2,spec}. They make use of a general rank-2 tensor with
symmetric and antisymmetric parts $e_{\mu\nu} = e_{(\mu\nu)} + e_{[\mu\nu]}$.
In \cite{rank2,spec} we have found two families of Lagrangians ${\cal L}(a_1)$
and ${\cal L}_{nFP}(c)$ in $D=4$ which differ from the usual FP theory
nontrivially, in the sense that, they can not be brought to the FP form by any
local field redefinition. They depend on the arbitrary real parameters $a_1$
and $c$ respectively.

In the next section we generalize the models ${\cal L}(a_1)$ and ${\cal
L}_{nFP}(c)$ as obtained in \cite{spec} in $D=4$ to arbitrary dimensions $D\geq
3$. The generalization of ${\cal L}_{nFP}(c)$ is less trivial if compared to
${\cal L}(a_1)$. The resulting Lagrangian explicitly depends on the space-time
dimension, see (\ref{nFP1}). At zero mass we show that ${\cal L}_{nFP}(c)$
contains the same particle content of the massless FP theory. In the special
case of $D=3$ this means that the massless theory has no particle content. We
use this fact to deduce via master action a fourth order dual theory similar to
the NMG. We show that there is still a mismatch of local symmetries between the
second and fourth order terms just like in the NMG case although both are now
Weyl invariant.

Regarding the family ${\cal L}(a_1)$  we study its massless limit which
contains an additional scalar field. We derive a range, see (\ref{26b}), for
the constant $a_1$ via unitarity of the massless theory.

 In section 4 we
draw our conclusions.

\section{``Spin-2'' particles in  $D\geq 3$ dimensions}

There are three families \cite{rank2} of second order models in $D=4$ which
describe a massive spin-2 particle via a rank-2 tensor $e_{\mu\nu}$, one of
them is the well known massive $FP$ theory. Those families are independent
models in the sense that they can not be interconnected by any local field
redefinition. Moreover they have different local symmetries in their massless
cases. However, it was demonstrated in \cite{spec} that they can be related
with the help of a decoupled and non dynamic extra field, for this reason we
call this kind of field a spectator. We start with the generalization of ${\cal
L}_{nFP}[e]$ to arbitrary dimensions. An important ingredient in our spectator
approach is the usual FP model whose coefficients are $D$ independent:

\be {\cal L}_{FP}[e_{\alpha\beta}] = -\frac 12
\p^{\mu}e^{(\alpha\beta)}\p_{\mu}e_{(\alpha\beta)} + \frac 14
\p^{\mu}e\p_{\mu}e + \left\lbrack \p^{\alpha}e_{(\alpha\beta)}-\frac 12
\p_{\beta} e \right\rbrack^2 -\frac{m^2}2 ( e_{\mu\nu}e^{\nu\mu} - e^2 ) \, .
\label{lfp} \ee

\no The $m=0$ limit of (\ref{lfp}) is the usual linearized Einstein-Hilbert
Lagrangian and describes massless spin-2 particles, it is invariant under
reparametrizations plus antisymmetric shifts $\delta \, e_{\mu\nu} = \p_{\nu}
\xi_{\mu} + \Lambda_{\mu\nu}$,  where $\Lambda_{\mu\nu} = - \Lambda_{\nu\mu} $.

The non Fierz-Pauli (nFP) model suggested in \cite{nfp} in $D=4$ deserves such
name due to a real parameter $c$ in the mass term which is given now by the
combination $(e_{\mu\nu}e^{\nu\mu}+c\, e^2)$. The absence of ghosts in the free
theory does not depend on the FP fine tuning $c=-1$. Explicitly, see
\cite{nfp}, we have in $D=4$:

\bea {\cal L}^{D=4}_{nFP}(c) &=& -\frac 12
\p^{\mu}e^{(\alpha\beta)}\p_{\mu}e_{(\alpha\beta)}  + \frac 16
\p^{\mu}e\p_{\mu}e + \left\lbrack
\p^{\alpha}e_{(\alpha\beta)}\right\rbrack^2 \nn \\
&-&   \frac 13 \left( \p^{\alpha}e_{\alpha\beta}\right)^2 -\frac 13
\p^{\alpha}e_{\alpha\beta}\p^{\beta}e- \frac{m^2}{2}(e_{\mu\nu}e^{\nu\mu} + c
\,  e^2). \label{lnfp}\eea

 Contrary to the Lagrangian (\ref{lfp}), we are going to verify that the
coefficients in the derivative terms of ${\cal L}^{D}_{nFP}(c)$ depend  on the
spacetime dimension. In order to generalize to $D$ dimensions we add a scalar
spectator $\varphi$ to the FP model following the same steps of \cite{spec},

\be {\cal L}_b = {\cal L}_{FP}[h_{\alpha\beta}] - b\, m^2 \, \varphi^2 +
h_{\alpha\beta}T^{\alpha\beta}, \label{lb} \ee

\no where ${\cal L}_{FP}[h_{\alpha\beta}]$ is the usual (symmetric) massive FP
theory given in (\ref{lfp}) with the replacement by a symmetric tensor
$e_{\alpha\beta}\to h_{\alpha\beta}=h_{\beta\alpha}$. We have also added a
symmetric external source $T^{\alpha\beta}$. The additional decoupled mass term
with arbitrary real constant $b$ clearly does not change the particle content
of the massive FP theory. After a generalized shift with  arbitrary real
constants $s$ and $t$:

\be h_{\mu\nu} \to h_{\mu\nu} +s\, \eta_{\mu\nu} \varphi +t\, \p_{\mu}\p_{\nu}
\varphi, \label{sh1} \ee

\no the Lagrangian ${\cal L}_b$ becomes

\bea {\cal L}_b &=& {\cal L}_{FP}[h_{\alpha\beta}]  + \left\lbrack s(D-2)-t\,m^2\right\rbrack(\p_{\mu}h\p^{\mu}\varphi-\p^{\mu}h_{\mu\nu}\p^{\nu}\varphi)+ \frac{m^2}{2}\left\lbrack s^2D(D-1)-2b\right\rbrack \, \varphi^2\nn\\
&+&
\left\lbrack \frac{s^2}{2}(D-2)(D-1)-s\,t\,m^2\,(D-1)\right\rbrack \p^{\mu}\varphi\p_{\mu}\varphi +s\, m^2\,(D-1) \varphi \, h\nn\\
&+& h_{\alpha\beta}T^{\alpha\beta} +s\, \varphi \, T +t\, \varphi \,
\p_{\mu}\p_{\nu} T^{\mu\nu}. \label{lb1} \eea

\no  By requiring that derivative couplings between $\varphi$ and $h$ vanish,
we fix $t=s(D-2)/m^2$. Introducing an auxiliary vector field and integrating by
parts  we can rewrite the $\p_{\mu}\varphi\p^{\mu}\varphi$ term in a first
order form

\bea {\cal L}_b &=& {\cal L}_{FP}[h_{\alpha\beta}]  +
\frac{m^2\,s^2}{2}\left\lbrack (D-1)(D-2)\right\rbrack A^{\mu}A_{\mu} +
\frac{m^2}{2}\left\lbrack s^2\,D(D-1)-2b\right\rbrack \, \varphi^2 +
h_{\mu\nu}T^{\mu\nu}
\nn\\
&-& s \, \varphi\left\lbrace\left\lbrack (D-1)(D-2)\right\rbrack \,s\, m \, \p
\cdot A - (D-1)\, m^2 h - T - \frac {(D-2)}{m^2}
\p_{\mu}\p_{\nu}T^{\mu\nu}\right\rbrace. \label{lb2} \eea


\no Due to the specific form of the usual Fierz-Pauli mass term, it is possible
to generate a Maxwell term by making another shift in ${\cal L}_b$ and using
the identity

\be {\cal L}_{FP}\left\lbrack h_{\mu\nu} + r\, (\p_{\mu}A_{\nu} +
\p_{\nu}A_{\mu})\right\rbrack = {\cal L}_{FP} [h_{\mu\nu}] - \frac{m\, r^2}2
F_{\mu\nu}^2(A) + 2\, m^2\, r\, A^{\mu}(\p^{\alpha}h_{\alpha\mu}-\p_{\mu}h).
\label{i2} \ee

\no After the shift $h_{\mu\nu}\to
h_{\mu\nu}+r(\p_{\mu}A_{\nu}+\p_{\nu}A_{\mu})$ in (\ref{lb2}) we decouple
$A_{\mu}$ and $\varphi$ by choosing $r=-s(D-2)/2m$. Next, we introduce an
antisymmetric field $B_{\mu\nu}$ by rewriting the Maxwell term in a first order
form, we end up with a master Lagrangian which now involves three extra fields
$(\varphi,A_{\mu},B_{\mu\nu})$ besides $h_{\mu\nu}$:

\bea {\cal L}_{M1} &=& {\cal L}_{FP}[h_{\mu\nu}] + \frac{m^2\,s^2}{2}(D-1)(D-2)
A^{\mu}A_{\mu} + \frac{m^2}{2}\left\lbrack s^2\,D(D-1)-2b\right\rbrack \, \varphi^2\nn\\
&+& (D-2)\, m\, s \, A^{\mu}\left( \p^{\alpha}B_{\alpha\mu} +
\p^{\alpha}h_{\alpha\mu} - \p_{\mu} h - \frac{\p^{\alpha}T_{\alpha\mu}}{m^2}
\right) \nn \\ &+& \frac{m^2}2 B_{\mu\nu}^2   + s \varphi \left\lbrack (D-1)\,
m^2 h + T + \frac{2}{m^2} \p_{\mu}\p_{\nu}T^{\mu\nu} \right\rbrack  +
h_{\mu\nu} T^{\mu\nu}. \label{lm1} \eea

We can define the generating function

\be Z_{M1}[T] = \int {\cal D}h_{\mu\nu} {\cal D}\varphi {\cal D}A_{\mu} {\cal
D}B_{\mu\nu} \,   e^{i \int d^Dx \, {\cal L}_{M1} }. \label{zm} \ee

\no If we functionally integrate over the extra field $B_{\mu\nu}$ in
(\ref{zm}) and reverse the  shift in (\ref{i2}), integrate over $A_{\mu}$ and
reverse the shift (\ref{sh1}) we come back to (\ref{lb}). On the other hand, if
we integrate over $\varphi$ and $A_{\mu}$  in first place we obtain the
Lagrangian \footnote{ We assume $s^2D(D-1)-2b\neq 0$ in order to integrate over
$\varphi$ in (\ref{lm1})}

%

\bea {\cal L}(s,b) &=& {\cal L}_{FP}[h_{\mu\nu}] + \frac{m^2}2 B_{\mu\nu}^2  -
\frac{(D-2)}{2(D-1)} \left( \p^{\alpha}B_{\alpha\mu} + \p^{\alpha}h_{\alpha\mu}
- \p_{\mu} h - \frac{\p^{\alpha}T_{\alpha\mu}}{m^2} \right)^2 \nn \\ &-&
\frac{s^2}{2\, m^2\left\lbrack s^2D(D-1) - 2b\right\rbrack}\left\lbrack (D-1)
\, m^2 \, h + T + \frac{(D-2)}{m^2}\p_{\mu}\p_{\nu}T^{\mu\nu} \right\rbrack^2.
\label{lsb} \eea

\no Defining $e_{\mu\nu} = h_{\mu\nu} + B_{\mu\nu} $, the Lagrangian ${\cal
L}(s,b)$ can be rewritten as

\be {\cal L}(s,b) = {\cal L}^{D}_{nFP}(c) + h^*_{\mu\nu}T^{\mu\nu} + {\cal
O}(T^2),  \label{lm3}\ee

\no where ${\cal O}(T^2)$ stands for quadratic terms in the source and:

\bea {\cal L}^{D}_{nFP}(c)&=& -\frac{1}{2}\partial_{\mu}e_{(\alpha\beta)}\partial^{\mu}e^{(\alpha\beta)}+\frac{1}{2(D-1)}\partial_{\mu}e[\partial^{\mu}e-2\partial_{\nu}e^{(\nu\mu)}]+[\partial_{\mu}e^{(\mu\nu)}]^{2}\nn\\
&-&\frac{(D-2)}{2(D-1)}[\partial_{\mu}e^{\mu\nu}]^{2}-\frac{m^{2}}{2}(e_{\mu\nu}e^{\nu\mu}+c\,
e^{2}).\label{nFP1}\eea

\no The linear term in the source in (\ref{lsb}) defines the dual field: \be
h^*_{\mu\nu} = e_{(\mu\nu)} - \frac{1+c}{(D-1)} \eta_{\mu\nu} e -
\frac{(1+c)(D-2)}{m^2\,(D-1)} \p_{\mu}\p_{\nu}e
-\frac{(D-2)}{2m^2\,(D-1)}\left( \p_{\mu}\p^{\alpha}e_{\alpha\nu} +
\p_{\nu}\p^{\alpha}e_{\alpha\mu}\right),  \label{hdual} \ee

The arbitrary parameter $c$ in the mass term of (\ref{nFP1})  is defined
through

\be c=\frac{2b-s^2(D-1)}{s^2D(D-1)-2b}. \label{bc} \ee

\no Since (\ref{lb}) and (\ref{lm3}) stem from the same generating function
(\ref{zm}) we conclude that ${\cal L}_{nFP}(c)$ and ${\cal L}_{FP}$ are dual to
each other in the sense that there is an equivalence of correlation functions
up to contact terms via dual map $h_{\mu\nu}^{*}\leftrightarrow h_{\mu\nu}$
i.e,

\be \langle
h^{*}_{\mu_1\nu_1}(x_1)...h_{\mu_N\nu_N}^{*}(x_N)\rangle_{nFP(c)}=\langle
h_{\mu_1\nu_1}(x_1)...h_{\mu_N\nu_N}(x_N)\rangle_{FP}+ contact\,\,\,
terms.\label{cts}\ee

\no It can be shown that the correlation functions involving the antisymmetric
tensor $B_{\mu\nu}$ in ${\cal L}_{nFP}(c)$ vanish identically up to contact
terms. Notice also that (\ref{nFP1}) reduces to (\ref{lnfp}) in $D=4$.

The equations of motion from (\ref{nFP1}) are given by:

\bea &&\square{e}^{(\mu\nu)}+\frac{1}{(D-1)}\partial^{\mu}\partial^{\nu}e-\partial^{\mu}\partial_{\lambda}e^{(\nu\lambda)}-\partial^{\nu}\partial_{\lambda}e^{(\lambda\mu)}-\frac{\eta^{\mu\nu}}{(D-1)}[\square{e}-\partial_{\alpha}\partial_{\beta}e^{\alpha\beta}]\nn\\
&&=m^{2}(e^{\nu\mu}+c\,\eta^{\mu\nu}e)-\frac{(D-2)}{(D-1)}\partial^{\mu}\partial_{\lambda}e^{\lambda\nu}.\label{nFP3}\eea

Applying $\eta_{\mu\nu}$ and $\partial_{\nu}$ on the equation (\ref{nFP3}), we
find $(cD+1)e=0$ and $\partial_{\mu}e^{\mu\nu}=-c\,\partial^{\nu}e$,
respectively. First, let us assume that $c\neq-1/D$. In this case the traceless
condition $e=0$ arises naturally as well as $\partial_{\mu}e^{\mu\nu}=0$. Then,
if we apply $\partial_{\mu}$ on the equation (\ref{nFP3}), using the fact that
$e_{\mu\nu}$ is traceless and transverse with respect to the first indice, we
obtain $\partial_{\mu}e^{\nu\mu}=0$. Coming back with these results in
(\ref{nFP3}) we can easily verify that $e_{[\mu\nu]}=0$ and rewrite the
equations of motion as a Klein-Gordon equation:

\be (\square-m^{2})e^{(\mu\nu)}=0.\label{nFP7}\ee

Therefore, we have a massive particle with $D(D-1)/2-1$ degrees of freedom for
arbitrary values of $c$. On the other hand if $c=-1/D$, then (\ref{nFP1})
becomes Weyl invariant, and we could fix the gauge $e=0$ and check that the
Fierz-Pauli conditions still remain satisfied. Anyway, the equations of motion
describe a massive ``spin-2'' particle in $D$ dimensions.

It is straightforward to check that for an arbitrary value of $D$, the massless
version of (\ref{nFP1}) is invariant under linearized reparametrizations plus
Weyl transformations, i.e, $\delta
e_{\mu\nu}=\p_{\nu}\xi_{\mu}+\eta_{\mu\nu}\phi$. It is possible to rewrite the
derivative terms in (\ref{nFP1}) in terms of the traceless tensor
$e_{\mu\nu}-\eta_{\mu\nu}e/D$ but the explicit dependence on $D$ does not
disappear.

We remark that if $b=0$, the arbitrary parameter $s$ disappears from
(\ref{lsb}) and we end up with an off-shell traceless description of massive
spin-2 particles  ${\cal L}_{nFP}(c=-1/D)$, see (\ref{bc}), confirming that the
arbitrariness of the ${\cal L}_{nFP}(c)$ family stems indeed from the arbitrary
mass term in (\ref{lb}) and not from the arbitrariness in the shift
(\ref{sh1}).

Now, it is a good moment to introduce the third family of Lagrangians
describing a massive spin-2 particle. This family depends on a real constant
$a_1$, see \cite{spec} for a detailed discussion. It is given by:

\bea {\cal L}_{a_1} &=& -\frac 12
\p^{\mu}e^{(\alpha\beta)}\p_{\mu}e_{(\alpha\beta)} + \left( a_1 + \frac 14
\right) \p^{\mu}e\left\lbrack \p_{\mu}e - 2
\p^{\alpha}e_{(\alpha\mu)}\right\rbrack + \left\lbrack
\p^{\alpha}e_{(\alpha\beta)}\right\rbrack^2 \nn \\
&+& \left( a_1 - \frac 14 \right) \left( \p^{\alpha}e_{\alpha\beta}\right)^2 -
\frac{m^2}{2}(e_{\mu\nu}e^{\nu\mu} - e^2) \quad . \label{la1} \eea

It can be shown that the equations of motion of ${\cal L}_{a_1}$ lead to the
Fierz-Pauli conditions $e_{[\mu\nu]}=0$, $\p^{\mu}e_{\mu\nu}=0$, $e=0$ and the
Klein-Gordon equation $(\Box-m^2)e^{(\mu\nu)}=0$ for arbitrary $D$. There are
three special values of $a_1$. First, if $a_1=-1/4$ the model (\ref{la1})
reduces to the one obtained in \cite{ms} via a different dualization procedure.
Secondly, if $a_1=1/4$, ${\cal{L}}_{a_1}$ reduces  to the FP theory
(\ref{lfp}). In the third case $a_1 = (3-D)/[4(D-1)]$, the Lagrangian
(\ref{la1}) becomes ${\cal L}_{nFP}(c=-1)$. So ${\cal L}(a_1)$ intersects both
previous families.

In its massless case ${\cal L}_{a_1}^{m=0}$ is invariant only under linearized
reparametrizations in general, $\delta \, e_{\mu\nu} = \p_{\nu} \xi_{\mu}$. In
the special cases $a_1=1/4$ and $a_1=(3-D)/\left\lbrack4(D-1)\right\rbrack$
 it is also invariant under $\delta
e_{\mu\nu}=\Lambda_{\mu\nu}$, where $\Lambda_{\mu\nu}=-\Lambda_{\nu\mu}$, and
$\delta e_{\mu\nu}=\phi\,\eta_{\mu\nu}$ respectively.

In order to check the particle content of ${\cal L}_{a_1}^{m=0}$ it is
convenient to  rewrite it with the help of a non-dynamical vector field
$v_{\mu}$ as follows

\bea
\mathcal{L}_{a_{1}}^{m=0}&=&-\frac{1}{2}\partial^{\mu}e^{(\alpha\beta)}\partial_{\mu}e_{(\alpha\beta)}+
\left( a_1 + \frac 14 \right) \p^{\mu}e\left\lbrack \p_{\mu}e - 2
\p^{\alpha}e_{(\alpha\mu)}\right\rbrack+\left\lbrack
\p^{\alpha}e_{(\alpha\beta)}\right\rbrack^2\nn\\
&&-\Big(a_{1}-\frac{1}{4}\Big)\left\lbrack
v_{\mu}v^{\mu}-2v_{\mu}(\partial_{\lambda}e^{(\lambda\mu)}+\partial_{\lambda}B^{\lambda\mu})\right\rbrack,\label{19}\eea

\no where we have used $e_{\mu\nu}=e_{(\mu\nu)}+B_{\mu\nu}$ with
$B_{\mu\nu}=-B_{\nu\mu}$. In first place, if we integrate over $v_{\mu}$ we go
back to ${\cal L}_{a_1}$. However if we functionally integrate over
$B_{\mu\nu}$ we have a constraint whose general solution is
$v_{\mu}=\partial_{\mu}\psi$, where $\psi$ is an arbitrary scalar field.
Substituting this result back in (\ref{19}) and changing variables
$\psi=\phi-e$ where $\phi$ is an arbitrary scalar field, we will find after a
field redefinition $e_{(\mu\nu)}\rightarrow
\tilde{e}_{(\mu\nu)}-2\frac{(a_{1}-1/4)}{(D-2)}\eta_{\mu\nu}\phi$, the
decoupled theory

\be \mathcal{L}_{a_{1}}^{m=0}=\mathcal{L}_{FP}^{m=0}[\tilde{e}_{\alpha\beta}]-
2\frac{(D-1)}{(D-2)}\Big(a_{1}-\frac{1}{4}\Big)\left\lbrack a_1 +
\frac{(D-3)}{4(D-1)} \right\rbrack
\partial_{\mu}\phi\partial^{\mu}\phi \quad .\label{mla1}\ee

\no where ${\cal L}_{FP}^{m=0}$ is given in (\ref{lfp})and corresponds to the
linearized version of the Einstein-Hilbert theory. Therefore, one can see that
the massless version of $\mathcal{L}_{a_{1}}$ describes a massless spin 2
particle plus a massless scalar field which disappears  at $a_1=1/4$ and
$a_1=-(D-3)/4(D-1)$. The scalar-tensor theory ${\cal L}_{a_1}^{m=0}$ is unitary
if: \be   a_1\leq - \frac{(D-3)}{4(D-1)}\quad;\quad a_1 \geq
\frac{1}{4}.\label{26b}\ee

\section{Fourth order massive spin-2 model in $D=3$}

In this section we check the particle content of the massless version of
(\ref{nFP1}) in arbitrary $D\geq 3$ and derive in $D=3$ a fourth order spin-2
model similar to the linearized NMG of \cite{bht}. In order to verify the
particle content of the massless version of (\ref{nFP1}), we introduce a
non-dynamical vector field $C_{\mu}$ and rewrite that Lagrangian as:

\bea \mathcal{L}_{nFP}^{D}(m=0)&=&-\frac{1}{2}\partial_{\mu}e_{(\alpha\beta)}\partial^{\mu}e^{(\alpha\beta)}+\frac{1}{2(D-1)}\partial_{\mu}e\partial^{\mu}e+\partial_{\mu}e^{(\mu\nu)}\partial^{\lambda}e_{(\lambda\nu)}-\frac{1}{(D-1)}\partial_{\mu}e\partial_{\nu}e^{(\nu\mu)}\nn\\
&&+\frac{(D-2)}{2(D-1)}C_{\mu}\left\lbrack
C^{\mu}+2\partial_{\nu}(e^{(\nu\mu)}+B^{\nu\mu})\right\rbrack.\eea

\no Integrating over $C_{\mu}$ one can recover the original kinetic term of
$\mathcal{L}_{nFP}^{D}$. Otherwise, the functional integration over
$B_{\mu\nu}$ in the path integral gives us a constraint whose general solution
is $C_{\mu}=\partial_{\mu}\phi$, where $\phi$ is an arbitrary scalar field.
Putting this back in $\mathcal{L}_{nFP}^{m=0}$ and making the change of
variables $e_{(\mu\nu)}\rightarrow
\tilde{e}_{(\mu\nu)}-(\tilde{e}+\phi)\eta_{\mu\nu}$ we get rid of $\phi$ and
obtain the linearized Einstein-Hilbert theory given in (\ref{lfp}) with $m=0$.
Namely,

\ \be \mathcal{L}_{nFP}^{m=0}\leftrightarrow {\cal
L}_{FP}^{m=0}\label{equi}.\ee

\no Thus $\mathcal{L}_{nFP}^{m=0}$ describes a massless particle with
$D(D-3)/2$ degrees of freedom which corresponds to a massless spin-2 particle
in $D=4$.

Since ${\cal L}_{nFP}^{m=0}$ has no \footnote{This is consistent with
(\ref{equi}) since the Einstein-Hilbert theory propagates no degrees of freedom
in $D=3$.} degrees of freedom in $D=3$, it can be used as a ``mixing term'' to
build up a master action \cite{dj} and deduce a higher order dual description
of massive spin-2 particles in $D=3$

We suggest the following master action in $D=3$:

\be
S_{M}=S_{nFP}^{D=3}[e_{\mu\nu}]-S_{nFP}^{m=0}[e_{\mu\nu}-f_{\mu\nu}],\label{master}\ee

\no where we have introduced another rank-2 tensor $f_{\mu\nu}$. Let us
introduce sources $j_{\mu\nu}$ which will allow us to derive a dual map between
correlation functions in the dual theories via the generating function:

\be W[j] =\int\, {\cal D}e_{\mu\nu}\, {\cal D}f_{\mu\nu} \, exp \,\,i\left(
S_{M} + \int d^3x\, j_{\mu\nu}e^{\nu\mu}\right)\label{wj}\ee

\no First of all, it is straightforward to verify that if we make the shift
$f_{\mu\nu}\to f_{\mu\nu}+e_{\mu\nu}$ in (\ref{master}) we decouple
$f_{\mu\nu}$ and we end up with the particle content of the massive action
$S_{nFP}^{D=3}[e]$ since $S_{nFP}^{m=0}$ has no content at all. So
(\ref{master}) certainly describes a massive spin-2 particle in $D=3$.

On the other hand, if we do not realize any shift and  integrate over the field
$e_{\mu\nu}$ we obtain the following higher order dual theory\footnote{The
theory ${\cal L}_{Weyl}$ has been obtained before in \cite{jm} via a
dimensional reduction of the massless FP theory in $D=4$.}  written in terms of
a traceless nonsymmetric tensor $\ttf_{\mu\nu}$

\be {\cal L}_{Weyl} = - {\cal L}_{nFP}^{m=0}[\ttf_{\mu\nu}] + {\cal
L}_K[\ttf_{\mu\nu}] + j_{\mu\nu}F^{\nu\mu}(\ttf) + {\cal O}(j^2) \quad ,
\label{lw1} \ee

\no where $ \ttf_{\mu\nu} = f_{\mu\nu} - \eta_{\mu\nu} \, f/3 $. The Lagrangian
${\cal L}_{nFP}^{m=0}$ is given in (\ref{nFP1}) with $D=3$ and $m=0$, while
${\cal L}_K = \left( R_{\mu\nu}^2 - \frac 38 R^2 \right)_{ff}/(2 \, m^2)$ is
the linearized version of the so called $K$-term of the New Massive Gravity of
\cite{bht} with $g_{\mu\nu} = \eta_{\mu\nu} + f_{(\mu\nu)} $. The nonsymmetric
tensor $F^{\mu\nu}(\ttf)$  plays the role of a dual field as we explain below.
It is given by the traceless combination

\be F^{\mu\nu}(\ttf)= \frac{1}{m^2}\left\lbrack  -\Box
\ttf^{(\mu\nu)}+\p^{\nu}\p_{\lambda}\ttf^{(\lambda\mu)}
+\frac{1}{2}\p^{\mu}\p_{\lambda}\ttf^{\nu\lambda}-\frac{\eta^{\mu\nu}}{2}
\p_{\alpha}\p_{\beta}\ttf^{\alpha\beta}\right\rbrack. \label{fdual} \ee

\no Dropping the sources, ${\cal L}_{Weyl}$ can be written as

\be {\cal L}_{Weyl} = \frac{\ttf_{(\mu\nu)}(\Box - m^2 )\ttf^{(\mu\nu)}}{2\,
m^2} + \frac{\p^{\mu}\ttf_{(\mu\nu )} (\Box - m^2
)\p_{\lambda}\ttf^{(\nu\lambda )}}{ m^2}
 + \frac{\left( \p_{\mu}\p_{\nu}\ttf^{\mu\nu}
\right)^2}{4\, m^2} + \frac{\left( \p_{\mu}\ttf^{\mu\nu} \right)^2}4 \quad .
\label{lw2} \ee

\no The Weyl symmetry is hidden in the definition of  $ \ttf_{\mu\nu}$. The Weyl theory is also invariant under
transverse linearized reparametrizations $\delta \ttf_{\mu\nu} = \p_{\nu} \zeta_{\mu} $, with $\p \cdot \zeta =
0$. The antisymmetric components $\ttf_{[\mu\nu]}$ only appear in the last term of (\ref{lw2}). The equations of
motion $K^{\mu\nu} = \delta\, S_{Weyl}/\delta\, \ttf_{\mu\nu} =0$ at vanishing sources can be written as

\bea K^{\mu\nu} &=& \frac{\left(\Box - m^2 \right)}{m^2}\left\lbrack \Box \,
\ttf^{(\mu\nu)} -  \p^{\mu}\p_{\alpha}f^{(\alpha\nu)} -
\p^{\nu}\p_{\alpha}f^{(\alpha\mu)} + \eta^{\mu\nu}
\frac{\p_{\alpha}\p_{\beta}\ttf^{\alpha\beta}}{2} \right\rbrack \nn \\ &+&
\frac 1{2\, m^2} \p^{\mu}\p^{\nu} \left( \p_{\alpha}\p_{\beta}
\ttf^{\alpha\beta} \right) - \frac 12 \p^{\mu}\p_{\alpha} \ttf^{\alpha\nu} = 0
\quad . \label{eomw} \eea

\no From the antisymmetric components $K^{[\mu\nu]}=0$ we have

\be \p_{\alpha}\ttf^{\alpha\nu} = \p^{\nu} \Phi \quad , \label{phi} \ee

\no where $\Phi$ is some scalar field. Using such information, the equations of
motion (\ref{eomw}) can be written as a Klein-Gordon equation for the dual
field (\ref{fdual}) :

\be K^{\mu\nu} = (\Box - m^2)F^{\mu\nu} = 0 \quad . \label{kg} \ee

\no Due to (\ref{phi}) we have $F^{[\mu\nu]} = 0 $. In summary, besides the
Klein-Gordon equation, all Fierz-Pauli conditions are satisfied by
$F^{\mu\nu}$,

\bea \eta_{\mu\nu}F^{\mu\nu} &=& 0 \quad , \label{fz1} \\
\p_{\mu}F^{\mu\nu} &=& 0 \quad , \label{fz2} \\
F^{[\mu\nu]} &=& 0 \quad , \label{fz3} \label{fpc} \eea

\no Thus, ${\cal L}_{Weyl}$ correctly describes a massive ``spin-2'' particle
in $D=3$.

It is typical of dual theories that equations of motion on one side may turn
into identities on the dual side. In the usual Fierz-Pauli theory written in
terms of a symmetric tensor $h_{\mu\nu}$, the traceless and transverse
conditions are dynamic while $h_{[\mu\nu]}=0$ is an identity. In the dual Weyl
theory $\eta_{\mu\nu}F^{\mu\nu}=0$ and $\p_{\nu}F^{\mu\nu}=0$ are identities
which do not depend on (\ref{phi}) as one can check directly from
(\ref{fdual}). The other $FP$ conditions $F_{[\mu\nu]}=0$ follow from the
equations of motion: $\delta S_{Weyl}/\delta \ttf_{\mu\nu}=0$.

One can go beyond the duality at classical level and obtain the quantum
equivalence between correlation functions by deriving with respect to the
source in (\ref{wj}) and (\ref{lw1}) obtaining:

\be \langle e_{\mu_1\nu_1}(x_1) ... e_{\mu_N\nu_N}(x_{N})\rangle_{nFP}=\langle
F_{\mu_1\nu_1}[\ttf(x_1)] ... F_{\mu_N\nu_N}[\ttf(x_{N})]\rangle_{Weyl} +\, \rm
{contact\,\, terms},\label{correlation}\ee

\no where the contact terms appear due to the quadratic terms in the sources in
(\ref{lw1}). In conclusion we have the dual map below between $S_{nFP}$ and
$S_{Weyl}$:

\be e_{\mu\nu} \leftrightarrow F_{\mu\nu}(\ttf)\quad.\label{dualmap1}\ee

Since the equations of motion in general are enforced at quantum level up to
contact terms \footnote{See the third footnote of \cite{spec}.}  we can use
(\ref{hdual}), (\ref{cts}), (\ref{correlation}) and the remark bellow
(\ref{cts}) to establish the direct dual map between the massive FP theory and
$S_{Weyl}$:

\be \langle e_{\mu_1\nu_1}(x_1) ... e_{\mu_N\nu_N}(x_{N})\rangle_{FP}=\langle
F_{\mu_1\nu_1}[\ttf(x_1)] ... F_{\mu_N\nu_N}[\ttf(x_{N})]\rangle_{Weyl} +\, \rm
{contact\,\, terms},\label{correlation2}\ee

\no Therefore, the equivalence between $S_{FP}$ and $S_{Weyl}$ holds true
beyond the on shell demonstration of \cite{jm}.

Notice that the identification (\ref{correlation2}) links gauge invariant
quantities since $F_{\mu\nu}[\ttf]$ is invariant under Weyl transformations and
transverse linearized reparametrizations while there is no local symmetry in
the $FP$ theory.

In summary, $S_{Weyl}$ describes a massive spin-2 particle  in $D=3$
(helicities $+2$ and $-2$). Although of fourth-order in derivatives the theory
is unitary, just like $S_{FP}$. Differently from $NMG$, both fourth and second
order terms can be written in terms of a traceless tensor (Weyl symmetry).
However , only the fourth order term is invariant under antisymmetric shifts
$\delta \ttf_{\mu\nu}=\Lambda_{\mu\nu}=-\Lambda_{\nu\mu}$. Consequently
$\ttf_{[\mu\nu]}$ is only present in the second order term.

\section{Conclusion}

Here we have shown that besides the paradigmatic Fierz-Pauli (FP) theory, there are other two families of second
order (in derivatives) Lagrangians describing massive ``spin-2'' particles in arbitrary $D\ge 3$ dimensions. The
new families require the use of a nonsymmetric second rank tensor ($e_{\mu\nu} \ne e_{\nu\mu}$). In particular,
one of the families is called non Fierz-Pauli (${\cal L}_{nFP}$) since the mass term does not need to fit in the
usual FP form, see (\ref{nFP1}). We have shown that at zero mass ${\cal L}_{nFP}(m=0)$ is equivalent to the
massless FP theory. Therefore, ${\cal L}_{nFP}^{D=3}(m=0)$ has no particle content. In a master action approach
\cite{dj}, Lagrangian terms with empty spectrum may be used to generate dual theories of higher order in
derivatives. In particular, this is how one can generate the $D=3$ new massive gravity (NMG) of \cite{bht} as a
dual theory to the usual (second order) massive FP theory, see comments in \cite{sd4}.

We have shown here by means of a master action that we can start with ${\cal
L}_{nFP}$ in $D=3$ and arrive at a dual theory describing a massive spin-2
particle in $D=3$ which contains a second order and a fourth order (in
derivatives) term just like the linearized new massive gravity, see
(\ref{lw1}). The fourth order term is the same one of the NMG, the so called
K-term. The K-term is invariant under linearized reparametrizations, linearized
Weyl transformations and antisymmetric shifts: $\delta e_{\mu\nu} = \p_{\mu}
\zeta_{\nu} + \p_{\nu} \zeta_{\mu} + \eta_{\mu\nu} \, \phi + \Lambda_{\mu\nu}
$, with $\Lambda_{\mu\nu}=-\Lambda_{\nu\mu}$. However, the second order term in
(\ref{lw2}) is not the linearized Einstein-Hilbert theory as in the NMG case,
the new second order term is invariant only under  $\delta e_{\mu\nu} =
\p_{\nu} \zeta_{\mu} + \eta_{\mu\nu} \, \phi $. The master action
(\ref{master}) has allowed us to prove that (\ref{lw1}) is off-shell equivalent
to the massive FP theory. The correlation functions of $e_{\mu\nu}(x)$ in the
massive nFP theory are mapped into correlation functions of $F_{\mu\nu}(f)$,
given in (\ref{dualmap1}), in the dual theory (\ref{lw1}) up to contact terms.

 The lack of Weyl symmetry in the second order term
(linearized Einstein-Hilbert) is a key obstacle for the renormalizability of
NMG \cite{deserprl,mo} since there will be a scalar degree of freedom
\cite{deserprl} which will be present only in the second order term. Contrary
to the linearized NMG, both terms of (\ref{lw1}) are invariant under Weyl
transformations. Unfortunately, it turns out that also in the case of
(\ref{lw1}), the second order term contains more degrees of freedom than the
fourth order one. The antisymmetric components $e_{[\mu\nu]}$ are not present
in the K-term which is invariant under antisymmetric shifts. Consequently the
propagator $\langle e_{[\mu\nu]}(p) \, e_{[\alpha\beta]}(-p)\rangle $ behaves
like $1/p^2$ for large momentum. So even if we were able to find a nonlinear
completion of the model (\ref{lw1}), the renormalizability of such model would
be jeopardized. It seems impossible,  using a rank-2 tensor, to formulate a
ghost free massive spin-2 model of higher order in derivatives where all
degrees of freedom are present in the highest order term. This might be a
signal that there is no renormalizable massive gravity even in $D=3$.

In section 2 we have shown that the second family of models ${\cal L}(a_1)$
found\footnote{The case $a_1=-1/4$ has been found before in \cite{ms}} in
\cite{rank2}, just like the usual Fierz-Pauli theory, has the same form in
arbitrary dimensions $D\ge 3$. At zero mass ${\cal L}(a_1)$ contains an
additional massless scalar particle besides the expected massless spin-2
particle unless $a_1=1/4$ or $a_1=-(D-3)/4(D-1)$ where ${\cal L}(a_1)$ reduces
to the ${\cal L}_{FP}$ and ${\cal L}_{nFP}(c=-1)$ respectively. The unitarity
bounds on the propagation of the massless scalar particle particle depends upon
the space-time dimension, see (\ref{26b}).

\section{Acknowledgements}

The work of D.D. is supported by CNPq (307278/2013-1) and FAPESP (2013/00653-4)
while A.L.R.S. is supported by Capes. We thank Gabriel B. de Gracia for a
discussion.

\end{document}